# Thermal Diffusivity and Viscosity of Suspensions of Disc Shaped Nanoparticles

Susheel S. Bhandari and K. Muralidhar

Department of Mechanical Engineering

Indian Institute of Technology Kanpur

Kanpur 208016 INDIA

and

Yogesh M. Joshi*

Department of Chemical Engineering

Indian Institute of Technology Kanpur

Kanpur 208016 INDIA

* Corresponding author, email: joshi@iitk.ac.in

## Abstract

In this work we conduct a transient heat conduction experiment with an aqueous suspension of nanoparticle disks of Laponite JS, a sol forming grade, using laser light interferometry. The image sequence in time is used to measure thermal diffusivity and thermal conductivity of the suspension. Imaging of the temperature distribution is facilitated by the dependence of refractive index of the suspension on temperature itself. We observe that with the addition of 4 volume % of nano-disks in water, thermal conductivity of the suspension increases by around 30%. A theoretical model for thermal conductivity of the suspension of anisotropic particles by Fricke as well as by Hamilton and Crosser explains the trend of data well. In turn, it estimates thermal conductivity of the Laponite nanoparticle itself, which is otherwise difficult to measure in a direct manner. We also measure viscosity of the nanoparticle suspension using a concentric cylinder rheometer. Measurements are seen to follow quite well, the theoretical relation for viscosity of suspensions of oblate particles that includes up to two particle interaction. This result rules out the presence of clusters of particles in the suspension. The effective viscosity and thermal diffusivity data show that the shape of the particle has a role in determining enhancement of thermophysical properties of the suspension.

## I. Introduction

Diffusion of momentum and heat in liquids containing suspended colloidal particles has significant academic importance and many technological applications. It is known that suspension of solid particles in liquid media enhances its viscosity (i.e. momentum diffusivity) and influences thermal diffusivity.[1] The changes in viscosity and thermal diffusivity of the suspension mainly depend on shape, and size of the suspended particles.[2-4] In addition, the nature of the surface such as charge, its distribution, and the presence of surfactant plays an important role.[5, 6] It has been noted that suspensions of anisotropic particles tend to show greater viscosity, namely, momentum diffusivity as well as enhanced thermal diffusivity, compared to isotropic, sphere-like, particles for a given volume fraction.[2, 3, 7-10] However, much less work has been carried out in the literature to study the behavior of suspensions of anisotropic particles in comparison to the isotropic. Specifically, among suspensions of anisotropic particles, very few studies investigate suspensions of disk-like (oblate) particles compared to those of rod-like (prolate) particles. Owing to greater demand for effective thermal transport from different kinds of devices whose length-scales range over six orders of magnitude (micrometers to meters), this is a continuing area of research.

The subject of heat transport through dilute suspensions of solid particles has received a renewed boost owing reports of enhancement in thermal conductivity of suspensions carrying nanoparticles.[8-11] This increase is well-beyond what is predicted by theoretical formulations that have been validated and known to work well for stable suspensions of larger particles. Such anomalous increase in thermal conductivity has been attributed by various authors to factors such as enhanced Brownian diffusivity,[12] phonon transport,[11] layering of liquid molecules over the particle surface,[13] and fractal cluster formation.[6, 14] Recent literature, however, indicates consensus on the possibility of cluster formation of nanoparticles as the factor responsible for the anomalous enhancement in thermal conductivity.[9] A major limitation in many studies that analyze enhancement of thermal conductivity in nanofluids by using theoretical models is unavailability of thermal conductivity of the nanoparticles. In such cases assumption of thermal conductivities over a certain range facilitates comparison of the trend. Interestingly, careful and independent investigations on this subject do not subscribe to the claim of anomalous enhancement of thermal conductivity beyond the theoretical.[15, 16] Poulikakos and coworkers[15] carefully studied the effect of particle size, concentration, method of stabilization and clustering on thermal transport in gold nanofluid. A maximum enhancement of 1.4 % was reported for 0.11 volume % suspension of 40 nm diameter particles in water suggesting no apparent anomaly.

In general, if thermal conductivity of the suspended particles is significantly greater than that of suspending liquid, increase in the aspect ratio is expected to cause substantial enhancement in thermal conductivity of the suspension. This is due to the fact that for the same volume fraction, anisotropy allows thermal transport over greater length-scales through elongated shape and greater surface area of the suspended particles. Here, the aspect ratio is defined in such a way that it is less than unity for oblate particles and greater than unity for the prolate. Based on Hamilton and Crosser theory,[3] Keblinski and coworkers[17] suggested that for equal volume fractions, a 10 fold enhancement in thermal conductivity is obtained when the aspect ratio of the anisotropic particle is around 100 (or 1/100). A good amount work has been carried out in the literature to investigate thermal properties of carbon nanotube suspensions in a variety of solvents.[8-11] Owing to different aspect ratios, solvents and type of nanotubes (single or multiwall), there is considerable spread in the data. The reported values range from 10 % to over 150 % enhancement in the thermal conductivity for 1 volume % concentration.[13, 18-23] On the other hand, experiments with aqueous suspension of titanate nanotubes having an aspect ratio of 10 is reported to show very moderate enhancement of around 3 % for 2.5 weight % suspension.[24] Recently Cherkasova and Shan[25] studied thermal conductivity behavior of suspension of whiskers having various aspect ratios (up to 10) and observed that for constant total volume fraction thermal conductivity increases with increase in the aspect ratio. The experimental data showed good agreement with effective medium theory of Nan and coworkers[26] which accounts for interfacial resistance in the conventional theories.

Compared to suspensions of rod-like particles, less work has been reported for suspensions of disk-like particles. Singh and coworkers [27] studied heat transfer behavior of aqueous suspension of silicon carbide having oblate shape and aspect ratio of around 1/4. They typically observed 30 % enhancement for 7 volume % concentration. The authors analyzed the data using Hamilton and Crosser theory[3] and found a correlation between them. Khandekar and coworkers[28] employed various spherical nanoparticle as well as Laponite JS (oblate shape, aspect ratio 1/25) based nanofluids in closed two phase thermosyphon and observed its heat transport behavior to be inferior to that of pure water in all cases. However, in a thermosyphon, in addition to thermal conductivity, fluid viscosity, wettability of the liquid on the surface of the apparatus, and roughness of the surface compared to particle dimensions also play an important role.[28] Recently, Bhandari and coworkers[29] reported transient heat transport behavior of a suspension of gel forming grade of nanoclay called Laponite RD, which has a soft solid-like consistency. Interestingly for about 1 volume % concentration of clay over 30 % enhancement in thermal conductivity was observed. Unlike the previous examples, this system is in solid state with finite elastic modulus (and infinite viscosity).[29] Recently many groups have studied heat transport in nanofluids composed of graphene. It has negligible thickness compared to its lateral dimensions and an extreme aspect ratio. Most studies employ water as the suspending medium. In order to render graphene hydrophilic, a functionalization step is carried out. For functionalized graphene – water nanofluids, thermal conductivity enhancement of 15 to 25 % is observed for around 0.05 volume %.[30-32] Other studies employ water soluble solvents such as ethylene glycol[33] and alcohol[34] and report similar enhancement as for water.

Although most studies on graphene nanofluid report thermal conductivity enhancement with increase temperature,[30-32, 35] one study reports it to remain constant.[34] Dhar et al.[36] and Martin-Gallego et al.[37] have analyzed thermal conductivity enhancement in graphene nanofluid and emphasized the importance of phonon transport. Two studies have compared the effect of particle shape, namely, prolate vis á vis oblate on the thermal conductivity enhancement at a given concentration using carbon nanotube and graphene nanofluids. Interestingly, Martin-Gallego et al.[37] observe enhancement in both the systems to be comparable while Wang and coworkers[35] report graphene nanofluid to perform better than nanotube suspension.

In the present work, we study transient heat transport behavior of an aqueous suspension of sol forming nanoclay - Laponite JS, which is an anisotropic, disk-shaped nanoparticle, using laser light interferometry.

## II. Viscosity and thermal conductivity of suspension of oblate spheroids

In the limit of very low volume fractions ($\phi$ < 0.03) an expression of viscosity of suspension having spherical particles is due to Einstein and is given by:[38]

$$\eta = \eta_s \left(1 + 2.5\phi\right)$$

where $\eta_s$ is viscosity of the Newtonian solvent in which the particles having volume fraction $\phi$ are suspended. This expression was obtained by estimating dissipation associated with flow around the sphere and assumes flow field around one sphere is not influenced by the presence of other spheres in the vicinity (therefore, the dilute limit). This expression was modified by incorporating hydrodynamic interactions associated with two–body interactions by Batchelor and Green.[39] The two body interaction leads to a contribution to $\eta$ proportional to $\phi^2$.

For a suspension containing monodispersed particles, the general expression for low shear viscosity irrespective of the aspect ratio of the particles can be written as:[40, 41]

$$\eta_r = \frac{\eta}{\eta_s} = 1 + C_1\phi + C_2\phi^2 + O(\phi^3), \tag{1}$$

where $C_1$ and $C_2$ respectively represent one and two body interactions and the volume fraction $\phi$. In this expression, $C_1 = [\eta]_0$ and $C_2 = k_H[\eta]_0^2$, where $[\eta]_0$ is an intrinsic viscosity in the limit of low shear rates. Parameter $k_H$ is the Huggins coefficient, which relates viscosity originating from two body interactions to that of viscosity associated with the infinite dilution limit where two body interactions are negligible. For particles having hard core interactions, the theoretical value of Huggins coefficient has been estimated to be $k_H \approx 1$ for spheres.[40] Theoretical value for disk-like particles is not available in the literature. Equation (1) with up to two body interaction ($\phi^2$ term) term holds good for $\phi \leq 0.1$.

In this work we study viscosity behavior of oblate spheroidal particles. We assume $d$ and $a$ to be the length-scales associated with short and long axis of the spheroid respectively. For a disk, $d$ and $a$ can be considered as a thickness and diameter respectively. Intrinsic viscosity $[\eta]_0$, a dimensionless quantity, is a function of the aspect ratio $r_p$ $(= d/a < 1)$ and is given by the expression:[41, 42]

$$[\eta]_0 = \frac{5}{2} + \frac{32}{15\pi r_p}(1 - r_p) - 0.628\left(\frac{1 - r_p}{1 - 0.075 r_p}\right). \qquad (2)$$

It can be seen that intrinsic viscosity expressed by Equation (2) is sensitive to the aspect ratio of the particle. It is inherently assumed in Equations (1) and (2) that particle orientation is isotropic, and therefore these equations are applicable only in the limit of small rotational Peclet number $(Pe << 1)$ defined as,[38]

$$Pe = \frac{4\eta_s d^3 \dot{\gamma}}{3kT}. \qquad (3)$$

In Equation (3), $\dot{\gamma}$ is shear rate, $k$ is Boltzmann constant, $\eta_s$ is solvent viscosity and $T$ is temperature.

The expression for electrical conductivity of a suspension containing oblate non-polarizable spheroidal particles is due to Fricke.[2] In this analysis, a single particle was placed in a unit cell, subjected to a unit potential difference, and analyzed in terms of Poisson's equation for a two-phase system. The influence of all other particles in the suspension on the near field of the single particle was taken to be equal to the mean value for the entire suspension. The prediction of electrical conductivity was seen to match very well with that of dog's blood for concentration of red corpuscles up to 90% in serum, the particles being treated as oblate spheroids. Treating thermal conductivity as analogous to electrical conductivity, the effective thermal conductivity of a suspension containing oblate spheroidal particles is given by:[2, 25]

$$k_e = k_s \left[ \frac{\zeta + (n-1) + (n-1)(\zeta - 1)\phi}{\zeta + (n-1) - (\zeta - 1)\phi} \right], \quad (4)$$

where $k_s$ is thermal conductivity of the solvent, $\zeta = k_p/k_s$ is particle-to-solvent conductivity ratio, and $k_p$ is thermal conductivity of the particles. Parameter $n$ is a shape factor given by:

$$n = \frac{\beta(\zeta - 1)}{(\zeta - 1) - \beta}, \quad (5)$$

Here $\beta$ is:

$$\beta = \frac{1}{3}\left[ \frac{4(\zeta - 1)}{2 + M(\zeta - 1)} + \frac{(\zeta - 1)}{1 + (1 - M)(\zeta - 1)} \right], \quad (6)$$

and

$$M = \frac{2\varphi - \sin 2\varphi}{2 \sin^3 \varphi} \cos \varphi, \quad (7)$$

where $\varphi = \cos^{-1}(r_p)$.

Rather than extending the expression derived for electric conductivity to thermal conductivity as carried out by Fricke,[2] Hamilton and Crosser[3] obtained effective thermal conductivity of a two component heterogeneous system by expressing it in terms of ratios of average temperature gradient in both the phases. They obtained the average gradient ratio by using expressions developed by Maxwell[7] and Fricke,[2] which led to the identical expression for $k_e$ as given by Equation (4) but a different shape factor:[3]

$$n = 3/\Psi, \tag{8}$$

where $\Psi$ is sphericity expressed as the ratio of surface area of a sphere having the same volume as the particle to that of surface area of the particle. For a disk having diameter $a$ and thickness $d$ (aspect ratio $r_p = d/a$), sphericity is given by the expression:

$$\Psi = \frac{2\left(3r_p/4\right)^{2/3}}{1 + r_p}. \tag{9}$$

The major difference between the analysis of Fricke and Hamilton and Crosser is the way effect of shape is incorporated. The latter considers the shape factor to be an empirical constant and expresses it as: $n = 3/\Psi$. On the other hand, in the Fricke expression shape factor is a function of thermal conductivities of particle as well as base fluid in addition to the aspect ratio. The expression by Hamilton and Crosser shows thermal conductivity of suspension to be slightly higher than that of Fricke for a given volume fraction of suspended particles. Both the expressions yield the Maxwell relation in the limit of spherical shape of the particle (aspect ratio of unity).

Similar to the expression for viscosity, a necessary condition for the application of Fricke model as well as Hamilton and Crosser model is an isotropic distribution of orientations of the disk-shaped particles. This requirement is ensured at the low Peclet number limit $Pe << 1$. In addition, sufficient time is needed to be given to the suspension in an experiment to stabilize so that Brownian motions randomize the orientation of clay discs and erase the shear history.

While the Fricke model is for electrical conductivity, it has been used to estimate thermal conductivity from the following thermodynamic perspective. Natural phenomena follow a cause-effect relationship that is expressible as in terms of forces and fluxes. It is a linear relation for electrical, thermal, hydraulic and other situations. Thus, one can have expressions for electrical, thermal, and hydraulic resistance and conductivity. Though the conductivities do not match in dimensional form, we can expect nondimensional conductivities of thermal and electrical systems scaled by a reference to match in quantitative terms.

**III. Determination of thermal diffusivity from an unsteady experiment**

Consider a horizontal layer of the aqueous suspension in one dimension of thickness $H$ maintained initially at constant temperature $T_C$. At time $t > 0$, the temperature of the top surface is suddenly raised to $T_H$ $(> T_C)$. Consequently heat diffuses from the top plate towards the bottom plate. The configuration is gravitationally stable in density. Hence, convection currents are absent and heat transfer is determined by pure diffusion. Combining Fourier's law of heat conduction with the first law of thermodynamics, the governing equation for temperature is given by,[1]

$$\frac{\partial \theta}{\partial \tau} = \frac{\partial^2 \theta}{\partial \psi^2}. \tag{10}$$

Here $\theta = (T - T_C)/(T_H - T_C)$ is dimensionless temperature, $\tau = \alpha t/H^2$ is dimensionless time, $\psi = y/H$ is a dimensionless coordinate measured from the lower surface. In addition, $t$ is time and $\alpha$ is thermal diffusivity of the suspension. The initial and boundary conditions associated with Equation (8) are:[29]

$$\begin{aligned} \theta &= 0 & \tau &\leq 0 \\ \theta &= 1 & \tau &> 0,\ \psi = 1 \\ \theta &= 0 & \tau &> 0,\ \psi = 0 \end{aligned} \tag{11}$$

Equation (10) can be analytically solved for a constant thermal diffusivity medium subject to conditions described by Equation (11) to yield:[29]

$$\theta = \psi - 2\sum_n \left[(-1)^{n+1}/n\pi\right] \sin(n\pi\psi) \exp(-n^2\pi^2\tau), \text{ for } n = 1, 2, 3 \ldots \tag{12}$$

Over the range of temperatures discussed in Section IV, thermal diffusivity changes by less than 0.5% and the constant diffusivity approximation is valid. According to Equation (12), the only material property that fixes the evolution of the temperature field is thermal diffusivity $\alpha$.

We describe in Section IV the experimental procedure to obtain temperature within the suspension as a function of position ($y$) and time. Using an inverse technique, we fit Equation (12) to the experimentally obtained temperature data so as to estimate thermal diffusivity. The technique relies on the method of least squares where the variance of the measured temperature relative to the analytical is minimized with respect to diffusivity as a parameter. In the limit of steady state ($\tau \to \infty$), Equation (12) becomes independent of thermal diffusivity. Hence, with increase in time, the fit of Equation (12) to the experimental data becomes progressively less sensitive to the choice of $\alpha$. In addition, with the approach of steady state, temperature field in the aqueous suspension is affected by the thermal properties of the confining surfaces. At the other extreme of small values of $\tau$, temperature gradients near the heated boundary are large and optical measurements are contaminated by refraction errors. Hence, with small time data, uncertainty associated with the estimated thermal diffusivity from Equation (12) is expected to be large. It can be concluded that the regression of experimental data against the analytical solution is most appropriate only at intermediate times. The conventional least squares procedure is thus modified to an equivalent weighted least squares procedure to highlight the data within a certain sensitivity interval, as discussed below.

Let $a_j$ $(j = 1, 2, ... m)$ parameters be determined from measurement data $Y_i$ $(i = 1, 2, ... n > m)$ using a mathematical model $T_i(a_j)$ by a least squares procedure. The process of minimizing the error functional:

$$E(a_j) = \sum_{i-1}^{n}(Y_i - T_i)^2 \qquad (13)$$

leads to the system of algebraic equations:[43]

$$\{a_j\} = [J^T J]^{-1} J^T \{Y_i\}. \qquad (14)$$

Here, $J$ is a Jacobian matrix defined as

$$J_{ij} = \frac{\partial T_i}{\partial a_j}. \qquad (15)$$

Equation (14) shows that uncertainty in the estimated parameters $a_j$ depends on the condition number of the matrix $J^T J$ since its inverse multiplies the measured data $Y_i$. Specifically, the condition number should be large for uncertainty to be small. This requirement leads to the result that the diagonal entries of the matrix $J^T J$ should be large. In this context, the Jacobian is referred to as sensitivity in the literature.[44]

During estimation of a single parameter, namely thermal diffusivity from spatio-temporal data of temperature, sensitivity estimates can be alternatively derived as follows. In analogy to Equation (15), the sensitivity parameter for the suspension as a whole is obtained by integrating over the layer thickness and is given by:[44]

$$S_t(t,\alpha) = \int_0^H \frac{\partial T}{\partial \alpha} dy. \qquad (16)$$

By differentiating Equation (12), the dimensionless form of the sensitivity parameter $\bar{S}_t$ is given by the expression:

$$\bar{S}_t = \frac{\alpha_R S_t}{H(T_H - T_C)} = -2\tau \sum_{n=1}^{\infty}\left[(-1)^{2n+1} + (-1)^{n+2}\right]\exp\left(-\bar{\alpha} n^2 \pi^2 \tau_R\right), \qquad (17)$$

where $\alpha_R$ is thermal diffusivity of a reference material (water in the present study), $\bar{\alpha} = \alpha/\alpha_R$, and $\tau_R = \alpha_R t/H^2$.

Analogous to time, a fit to the experimental data closer to either of the boundaries in not sensitive to the choice of $\alpha$. The loss of sensitivity arises from the fact that the walls are at constant temperature, independent of the material diffusivity. Following Equation (16), sensitivity associated with the experimental data as a function of the spatial coordinate can be estimated by integrating with respect to time and is given by:

$$S_y(y,\alpha) = \int_0^{\infty} \frac{\partial T}{\partial \alpha} dt. \qquad (18)$$

However, in Equation (18) rather than integrating $\partial T/\partial \alpha$ over the entire duration of experiment, we integrate the same over only that duration where sensitivity estimated from equations (16) and (17) is very high (typically above 90 % of the maximum in sensitivity). If interval between $t_1$ and $t_2$ represents that window of high sensitivity, Equation (18) can be modified to:

$$S_y(y,\alpha) = \int_{t_1}^{t_2} \frac{\partial T}{\partial \alpha} dt. \qquad (19)$$

In a manner similar to Equation (17), the dimensionless sensitivity parameter $\bar{S}_y$ for spatial sensitivity of data is given by:

$$\bar{S}_y = \frac{S_y \alpha_R}{(T_H - T_C)(t_2 - t_1)} = \frac{1}{\bar{\alpha}(\tau_{R2} - \tau_{R1})} \sum_n \frac{2}{n\pi} (-1)^{n+2} \sin(n\pi\psi) A_n, \tag{20}$$

where $A_n$ is given by:

$$A_n = \left(\tau_{R2} + \frac{1}{\bar{\alpha} n^2 \pi^2}\right) \exp\left(-\bar{\alpha} n^2 \tau_{R2}\right) - \left(\tau_{R1} + \frac{1}{\bar{\alpha} n^2 \pi^2}\right) \exp\left(-\bar{\alpha} n^2 \tau_{R1}\right). \tag{21}$$

In order to obtain a dependable value of $\alpha$, Equation (12) is fitted to the experimental data in that window of $\psi$ and $\tau$, where sensitivity, defined by Equations (17) and (20) is high. In addition, uncertainty in the estimated properties is determined by conducting statistically independent experiments.

**IV. Materials, sample preparation and experimental protocol**

In the present study, we use an aqueous suspension of Laponite JS® (sodium magnesium fluorosilicate), which is a sol forming variant of Laponite clay.[45] Laponite particles are disk-like with diameter 25 nm and thickness 1 nm with very little size/shape distribution.[46] In a single layer of Laponite, two tetrahedral silica layers sandwich an octahedral magnesia layer.[47] A particle of Laponite can, therefore, be considered as a single crystal. Face of Laponite particle is negatively charged while the edge has weak positive charge. Laponite JS used in the present work was procured from Southern Clay Products Inc. Laponite JS contains premixed tetrasodium pyrophosphate which prevents aggregation of the particles thereby providing stability to the sol. It should be noted that in our previous communication,[29] we had investigated gel forming grade: Laponite RD, which does not have tetrasodium pyrophosphate and therefore transforms into a soft solid with infinite viscosity and finite modulus. The suspension of Laponite JS used in this work, on the other hand, is a free flowing liquid with viscosity comparable to water. According to the manufacturer, suspension of Laponite JS is stable against aggregation over duration of months for much larger concentrations than used in the present work.[45] Predetermined quantity of white powder of Laponite was mixed with ultrapure water using an ultra Turrex drive as stirrer. Water was maintained at pH=10 to ensure chemical stability of the Laponite particles.[48] Laponite JS forms a stable and clear suspension in water that remains unaltered for several months. Laponite JS suspension prepared using this procedure was used for rheology and interferometry experiments. The ionic conductivity and pH of the suspension after incorporation of Laponite JS are reported in Table 1. While pH remains practically constant, conductivity increases as expected with Laponite concentration.

Rheological experiments were carried out in a concentric cylinder geometry (diameter of inner cylinder 26.663 mm and thickness of the annular region 1.08 mm) using Anton Paar MCR 501 rheometer (stress controlled rheometer with minimum torque of 0.02±0.001 µNm). Viscosity is measured over a range of shear rates 25/s to 140/s and is observed to be constant over this range. We ensured absence of Taylor-Couette instability in the annular gap during the measurement of viscosity over this shear rate range.

Transient heat transfer experiments were performed using a Mach-Zehnder interferometer. A schematic drawing of the setup is shown in Figure 1. In this setup the collimated laser beam [632.8 nm, 35 mW He-Ne laser (Spectra Physics)] is split using a beam splitter and passed through the test and the reference chambers as shown in Figure 1. Subsequently, the beams are superimposed, and the resulting image is recorded by a CCD camera. Owing to the optical path difference generated between the beams passing through the test and the reference sections an interference pattern is formed. Before beginning the experiment, the test chamber was filled with Laponite JS suspension having a certain concentration. The reference chamber was filled with glucose solution. The concentration of the glucose solution in the reference chamber was varied in such a way that it matched the refractive index of Laponite suspension in the test chamber at temperature $T_C$. This step ensured balancing of the optical path lengths of the light beams passing through the reference and the test chambers. Superposition of the two light beams resulted in constructive interference with a bright spot on the screen. This state of the interferometer is referred as the infinite fringe setting. Subsequent to attainment of thermal equilibrium between the test and the reference sections of the apparatus, the temperature of the top plate was raised to $T_H$. Soon after the top plate temperature is raised to $T_H$, heat diffuses towards the lower surface, causing an increase in temperature within the medium. Consequently, refractive index of the suspension is altered, generating an optical path difference between the beams passing through the test and the reference chambers. Superposition of the two light beams leads to an interference pattern.

In the present set of experiments, we maintained $T_C$ between 19 and 20°C, while $T_H$ is varied between 21 and 23°C. The horizontal bounding surfaces of the test cell were made of 1.6mm copper sheets. Temperatures were maintained constant by circulating water over these surfaces from constant temperature baths. The lower surface was baffled to form a tortuous flow path in addition to creating a fin effect. Constant temperature baths maintained temperature constant to within ±0.1°C. Direct measurement by thermocouples did not reveal any (measurable) spatial temperature variation. These temperatures were constant to within ±0.1°C during the experimental duration of 4-5 hours. Experiments were carried out in an air-conditioned room where temperature was constant to within ±1.0°C. The bounding surfaces reached their respective temperatures in 1-2 minutes. This delay is negligible in comparison to the experimental duration of 4-5 hours. The time constant of the diffusion process can be estimated as $H^2/\alpha \sim 4.6$ hours.

We study below the effect of concentration of Laponite JS up to 4 volume % in water. For each concentration, a minimum of three experiments were carried out to examine repeatability. In order to analyze results of laser interferometry, the temperature dependence of refractive index of the Laponite suspension at each concentration is required. In the present work, a refractometer (Abbemat 500, Anton Paar) was used to measure refractive index of the suspension at various temperatures.

**V. Results and Discussion**

Viscosity of the suspension

In Figure 2 we plot effective viscosity of the aqueous suspension as a function of volume fraction of Laponite JS. It can be seen that viscosity, shown as circles, increases with volume fraction, such that for 4% volume fraction of Laponite, viscosity is three times that of water. Note that the uncertainty band is quite small. In Section II we discussed the need to enforce a limit of $Pe << 1$ to ensure random (isotropic) orientation of the clay disks for Equations (1) to hold. This assumption can now be examined. For $\eta_s$=10$^{-3}$ Pa-s, $d$ = 25 nm, $\dot{\gamma}$ =100/s, and $T$ =300 K, we get $Pe \approx$ 0.0005, which is considerably smaller than unity. Hence, the particle orientation is randomized and there are no preferred directions for thermal diffusivity. Further, the system is gravitationally stable (bottom heavy) and convection is not present. Hence, random orientation of the original medium is preserved during the conduction heat transfer experiment.

It can be seen that Equation (1) fits the data very well, which is represented by a solid line. In the inset we plot $(\eta - \eta_s)/(\eta_s \phi_L)$ as a function of concentration of Laponite. The linear fit to the data in the inset in rearranged Equation (1) is given by: $(\eta - \eta_s)/(\eta_s \phi_L) = [\eta]_0 + C_2 \phi$, whose intercept on the vertical axis lead to intrinsic viscosity $[\eta]_0$. Equation (2) relates intrinsic viscosity $[\eta]_0$ to the aspect ratio of oblate particles. Remarkably value of $[\eta]_0$ =18.62 obtained from linear fit leads to $1/r_p$ =25.7 which is very close to the value of 25 reported in the literature from independent measurements.[45, 46] The fit of the data also yields Huggins coefficient $k_H$ =2 for disk like particles of Laponite. Importantly, a good fit of the experimental data is obtained to the viscosity relation based on classical theory. It confirms the aspect ratio of the particle independently and rules out the presence of clusters of disk-like particles within the suspension. For this reason, measured thermal conductivity data in the following section have been compared with well-established models of HC and Fricke.

Thermal diffusivity of the suspension

Next, we discuss thermal diffusivity measurements based on interferometry experiments. Subsequent to a step increase in temperature of the top surface, one dimensional diffusion of heat is initiated towards the cold plate of the test section. An increase in temperature of the suspension changes its density, and hence the refractive index. This alters the optical path length of the beam of light passing through the test section in comparison with the reference, producing an interference pattern on superposition. It is noted here that diffusivity changes in the range 21–23°C are small enough from a modeling perspective (Equation 12). However, density and refractive changes are large enough to create a measurable optical path difference needed for the formation of interferograms. Measurement sensitivity is further enhanced by having a test cell that is long in the viewing direction.

The time evolution of the interference pattern is shown in Figure 3. In the beginning, the fringe pattern at time $t=0$ reveals an infinite fringe setting, corresponding to constructive interference when equal path lengths are obtained in the light beam passing through test and reference sections. As heat diffuses from the top surface into the suspension, fringes appear in its neighborhood, and migrate with time towards the lower surface. In the limit of large times, steady state sets in and fringes get distributed throughout the field. The temperature difference associated with two consecutive fringes ($\Delta T_\varepsilon$) can be estimated from the knowledge of variation of refractive index with temperature $dn/dT$, wavelength of light $\lambda$, length of the test cell $L$, and is given by principles of wave optics as:[49]

$$\Delta T_\varepsilon = \lambda / (L\, dn/dT). \tag{22}$$

Equation (22) assumes that the passage of the light ray is straight and beam bending due to refraction is small. Refraction would be strong in the initial stages when temperature (and hence, refractive index) gradient near the top wall is large. Explicit calculations with a generalized form[49] of Equation 22 showed that for times greater than 20 minutes, refraction error is less than 1% and Equation (22) is applicable. This requirement is fulfilled during data analysis in a statistical framework using the sensitivity function. In Figure (4), we plot evolution of temperature as a function of distance from the top surface and time elapsed in minutes, since the creation of a temperature jump. Prior to the introduction of the jump in temperature, the temperature field is uniform, and equal to the temperature of the lower surface. As heat diffuses through the suspension, a sharp gradient in temperature develops near the top plate. The gradient weakens progressively, and in the limit of long time the steady state sets in.

The evolution of temperature showed in Figure 4 is equivalent to analytical solution of diffusion equation given by Equation (12). A fit of the analytical solution to the experimental data leads to the determination of thermal diffusivity $\alpha$. However, as discussed in Section II, the fit of experimental data in the limit of very short times and very long times is not sensitive to the parametric changes in $\alpha$ and can result in curve-fitting errors. In addition, temperature data close to the isothermal boundaries of the apparatus are also not sensitive to $\alpha$. Therefore, in order to obtain the appropriate range of times ($t$ or $\tau$) and distance from the lower surface ($y$ or $\psi$) to be considered for parameter estimation, we plot $\bar{S}_t$ as a function of $\tau$ and $\bar{S}_y$ as a function of $\psi$ in Figures 5(a) and 5(b) respectively. It can be seen that both sensitivities have a clear maximum, indicating that experimental data should be selectively used for parameter estimation. In Figure 4, we fit Equation (12) to only that part of the experimental data which is within the 90 % window of the maximum value of each of the sensitivities. This statistical procedure reliably estimates thermal diffusivity as a function of volume fraction of Laponite JS.

Thermal conductivity of the suspension

Thermal conductivity enhancement in nanofluids has been a subject of many studies over the past decade. Apart from solvent selection, types of nanoparticles, their sizes, and to a lesser extent, their shape have been explored. The measurement technique – point, line or volumetric, influences the conductivity prediction. It is, therefore, not surprising that contradictory opinions have been expressed in the literature. In this context, knowledge of physicochemical behavior of colloidal suspensions is also necessary. A rigorous study that employs thermal conductivity measurement in a stable suspension of anisotropic particles and comparison with available theories helps in further understanding of the subject.

We obtain thermal conductivity of the suspension using fitted values of thermal diffusivity. The two are related as $k = \rho C_P \alpha$, where $\rho$ is density of suspension while $C_P$ is its heat capacity at constant pressure. Heat capacity of Laponite clay is known to be 1.03 kJ/(kg-K),[38-39] while that of water is 4.18 kJ/(kg-K). Heat capacity of the suspension is estimated by using the rule of mixtures:[50, 51]

$$C_{P,sus} = w_L C_{P,L} + w_W C_{P,W}, \qquad (23)$$

where $w$ is mass fraction, $C_P$ is heat capacity at constant pressure, and subscripts $L$ and $W$ represent Laponite and water respectively. In Figure 6 we plot thermal conductivity of the suspension normalized by thermal conductivity of water as a function of volume fraction of Laponite. It can be seen that thermal conductivity increases with increase in concentration of Laponite. An increase of around 30 % is observed for a volume fraction of 4 %.

In Section II, we describe models for thermal conductivity proposed by Fricke[2] and Hamilton and Crosser[3] for a Brownian suspension of disk-like particles. Accordingly, the effective thermal conductivity of the suspension is represented by Equations (4) to (7) and Equations (4), (8) and (9) respectively. In order to estimate effective thermal conductivity, thermal conductivity of a single particle of Laponite is necessary, which is not available in the literature. In addition, owing to its nanoscopic size and anisotropic shape, it is expected that thermal conductivity of the Laponite crystal would be difficult to measure using conventional techniques. Owing to a layered structure of the particle, comprising outer silica layers that sandwich magnesia layer, a particle of Laponite is expected to have a high thermal conductivity.

In Figure 6, we plot a fit of Fricke model (Equations 4 to 7) and Hamilton and Crosser model (Equations 4, 8 and 9) to the experimental data. It can be seen that model shows a good fit to the suspension thermal conductivity for $k_p/k_f$ =13. Since thermal conductivity of water is 0.58 W/(m-K), the estimated thermal conductivity of the Laponite particle comes out to be around 7.54 W/(m-K). Interestingly crystalline silica is known to have thermal conductivity of 14 W/(m-K).[52] This suggests that the estimated value of thermal conductivity of Laponite particle is plausible. Figure 6 also shows prediction of Maxwell model (Equation 4) for suspension of isotropic particles ($n$=3) for $k_p/k_f$ =13. It can be seen that, owing anisotropy, Laponite JS suspension substantial improvement over suspension of isotropic particles that obey Maxwellian behavior. As discussed in previously such enhancement can be attributed to increased thermal transport due to greater length-scale and surface area associated with anisotropic particle than that of isotropic particles.

As discussed in Section I on introduction to the present study, a significant amount of research in the literature report the effect of incorporation of nanoparticles in liquid media on thermal conductivity. Some studies claim an anomalous increase in thermal conductivity that cannot be explained by the existing effective medium theories. It is now believed that an anomalous enhancement can be attributed to formation of a cluster with fractal-like behavior. In the present work, an excellent fit of the viscosity relation based on two body interaction to the experimental data rules out the possibility of cluster formation.

The thermal conductivity data based on unsteady heat conduction is less conclusive. The degree of enhancement of thermal conductivity of the aqueous suspension of Laponite and its match with theory relies on the knowledge of the particle thermal conductivity. The possibility of an anomalous increase can be ruled out only with an independent estimation of thermal conductivity of the Laponite particle. Nonetheless, qualitative agreement of the Fricke model and Hamilton and Crosser model with experimental data is encouraging. In view of the fact that the estimated value of Laponite thermal conductivity is plausible, reference to any other mechanism to support anomalous enhancement is not necessary. Following the correct prediction of viscosity, the effective medium theory may now be taken as applicable in the context of energy transfer. Consequently, the theoretical models combined with the measured thermal diffusivity data provide a method of estimating the particle thermal conductivity itself.

Thermal fluctuations

The role of Brownian motion on the enhancement of thermal conductivity has been assessed thoroughly in the literature. Careful analysis by Fan and Wang[9] (and references therein) clearly rule out this possibility. On thermophoresis, Piazza and Parola[53] discuss thermophoretic diffusivity to be comparable (or smaller) than mass diffusivity of Laponite in water. The mass diffusivity itself is several orders of magnitude smaller than thermal diffusivity. In addition the gradient of temperature in the thermal cell used in the present work is small, being around 2K at 300K over a distance of 50mm. These factors suggest that thermophoresisis has negligible contribution to the overall heat transport.

Anisotropic particles at the nanoscale have significant advantages over microscopic particles. Owing to their small size, thermal fluctuations of internal energy eliminate chances of sedimentation and enhance stability of the suspension. Furthermore, thermal factors keep small anisotropic particles isotropically orientated, making thermal conductivity an isotropic quantity. Larger anisotropic particles, on the other hand, would take longer to randomize their orientation.

## VI. Conclusions

We measure viscosity of an aqueous suspension of Laponite JS and observe that it increases with particle concentration. Effective medium theory for viscosity of suspensions of anisotropic particles fit the experimental data very well. In addition intrinsic viscosity estimated from the fit correctly predicts the aspect ratio of the Laponite particles. This suggests absence of formation of any particle clusters in the suspension. We measure thermal diffusivity from a transient heat conduction experiment using laser light interferometry, wherein temperature dependence of refractive index of the suspension is utilized for image formation. Typically for incorporation of 4 volume % of nanodisks, around 30 % enhancement in thermal conductivity is observed in the temperature range of 21-23°C. Similar enhancements can be expected at other temperatures that are close to the ambient. The Fricke model as well as the Hamilton and Crosser model for thermal conductivity of suspension of anisotropic particles explain the trend of the experimental data quite well. The fit requires information of particle thermal conductivity. A retrieved value of the Laponite thermal conductivity is seen to be realistic.

**Acknowledgement:** This work was supported by Department of Science and Technology, Government of India.

**Table 1.** Ionic conductivity and pH of Laponite JS suspension

|  | Initial water+NaOH (10pH) solution | Concentration of Laponite JS (Volume %) | | | | |
| --- | --- | --- | --- | --- | --- | --- |
|  |  | 0.6 | 1 | 1.4 | 2.4 | 4 |
| pH | 10 | 9.996 | 9.983 | 9.917 | 9.993 | 10 |
| Conductivity (mS/cm) | $2.911 \times 10^{-2}$ | 1.933 | 2.995 | 3.738 | 6.205 | 8.984 |

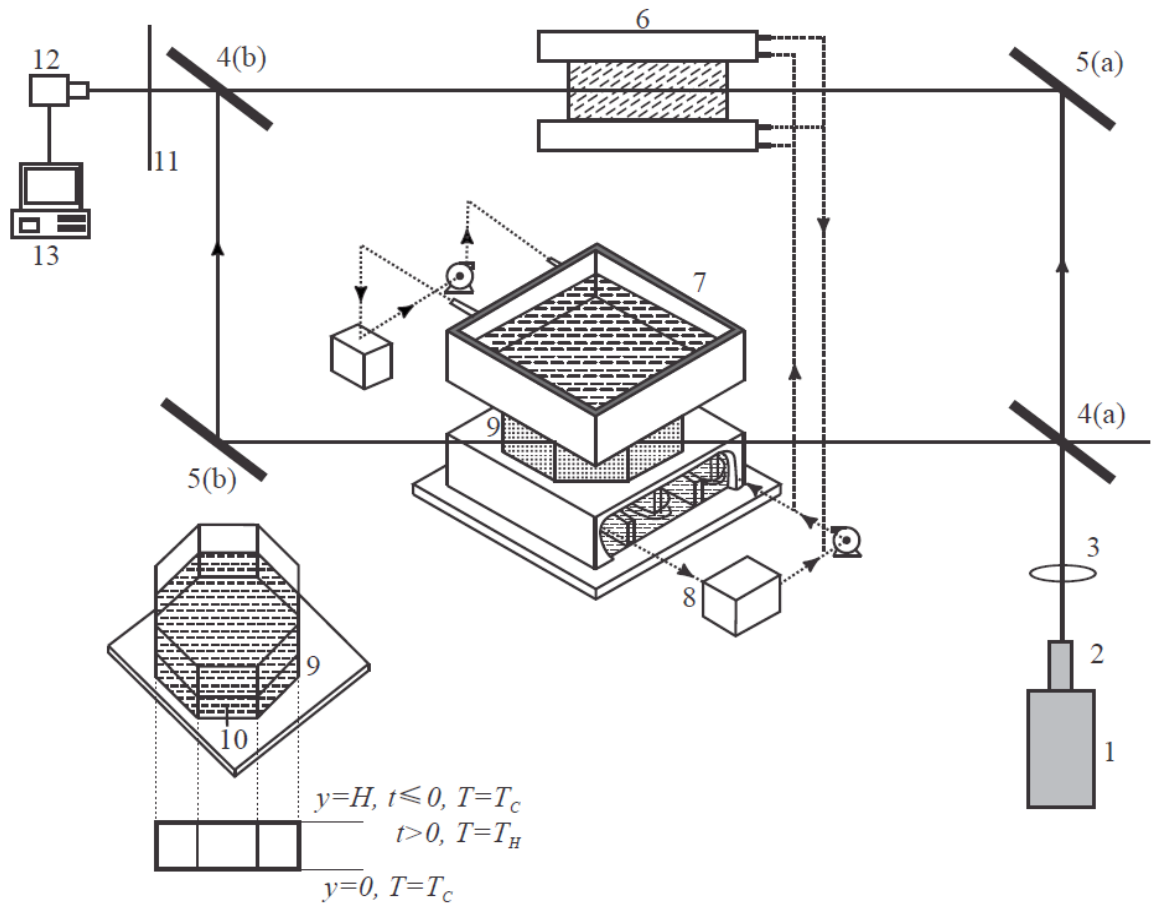

**Figure 1.** Schematic Drawing of a Differentially Heated Test Cell Integrated with the Mach-Zehnder Interferometer. 1, Laser; 2, Spatial Filter; 3, Plano Convex Lens; 4, Beam Splitters 4(a), 4(b); 5, Mirrors 5(a), 5(b); 6, Compensation Chamber; 7, Hot Water Bath; 8, Cold Water Bath; 9, Optical Cavity; 10, Optical Window; 11, Screen; 12, CCD Camera; 13, Computer.

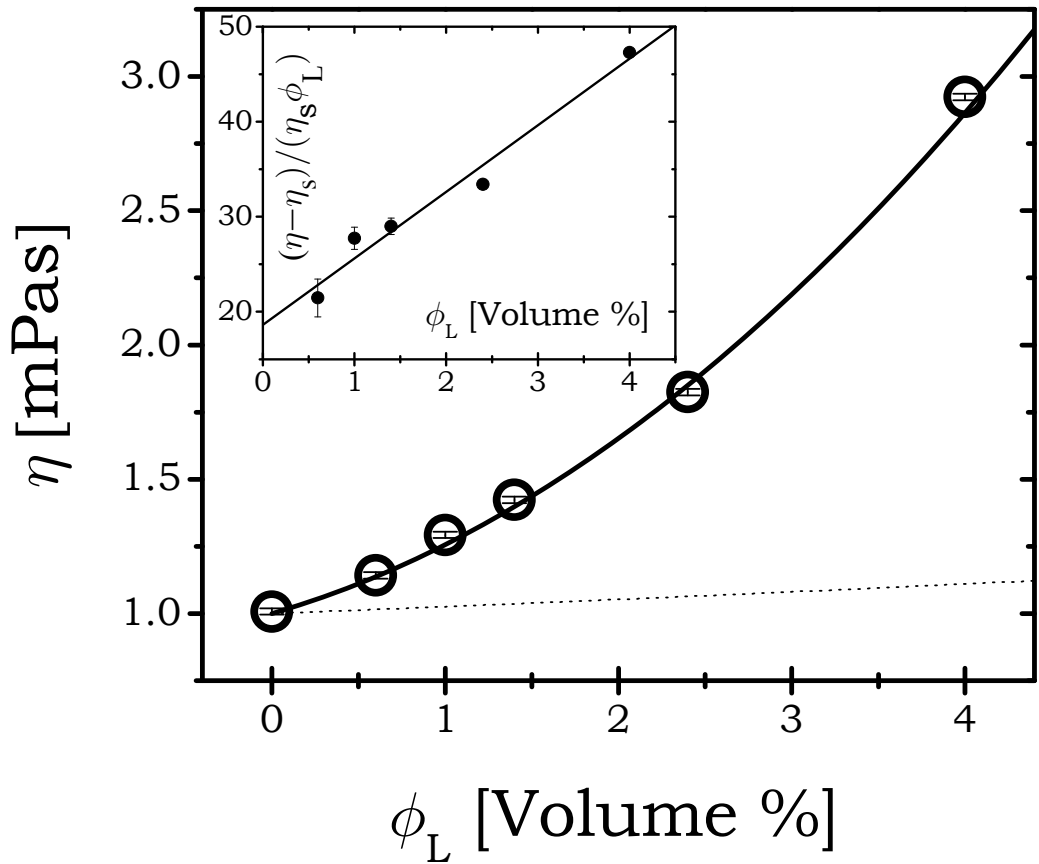

**Figure 2.** Viscosity (open circles) of suspension is plotted as a function of volume % of Laponite JS in water obtained at 25°C. Solid line is a fit of Equation (1) for oblate particle suspension. Dotted line is prediction of viscosity of suspension having spherical particles. In the inset $(\eta - \eta_s)/(\eta_s \phi_L)$ is plotted as a function of volume % of Laponite. The solid line in the inset is the same fit of Equation (1) shown in the main figure. The intercept of the line on the vertical axis (limit of $\phi_L \to 0$) is intrinsic viscosity.

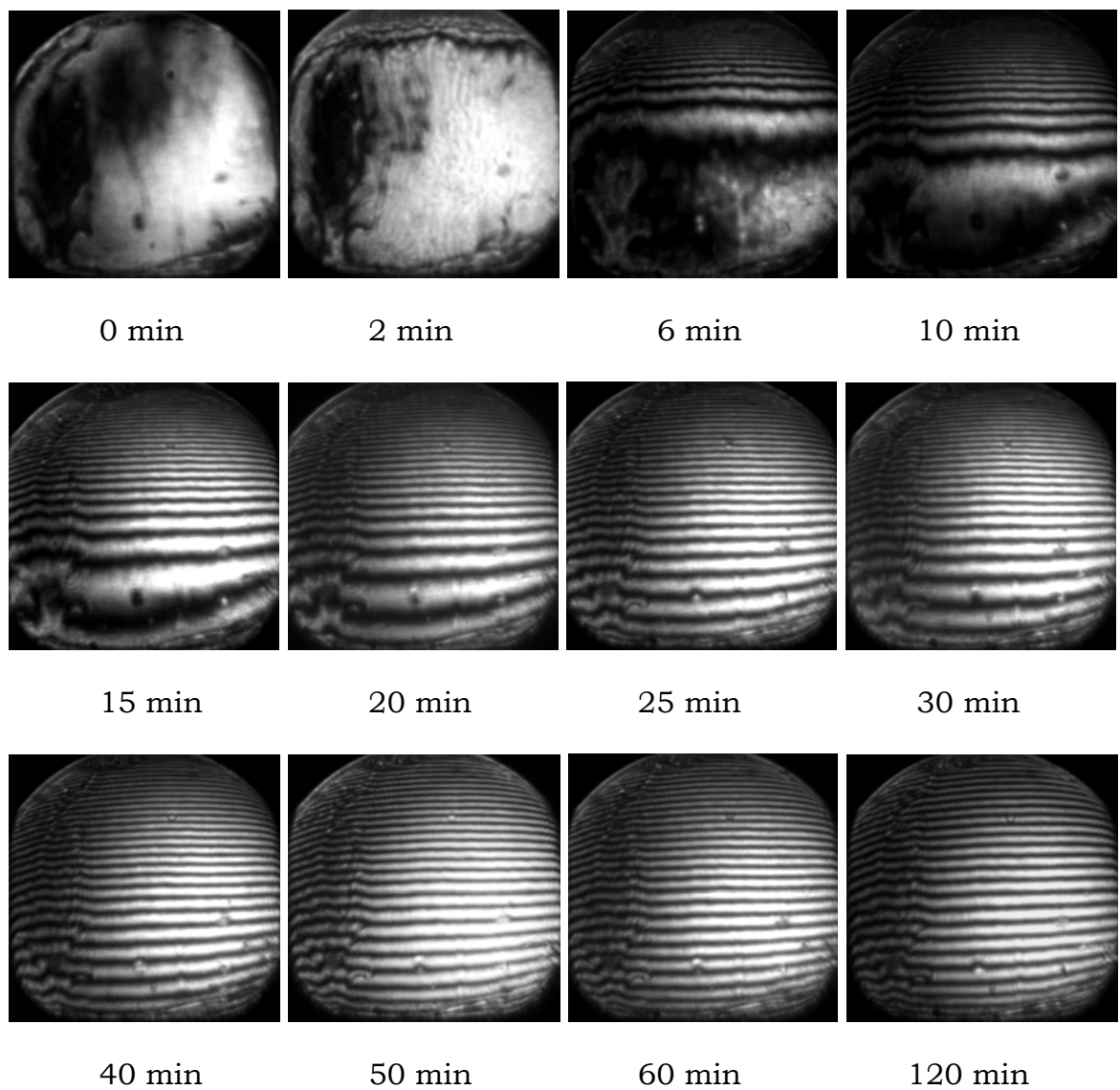

**Figure 3:** Evolution of interference pattern as a function of time for 2.4 volume % aqueous Laponite JS suspension. The upper and lower surfaces were maintained at 21.7°C and 19.7°C respectively.

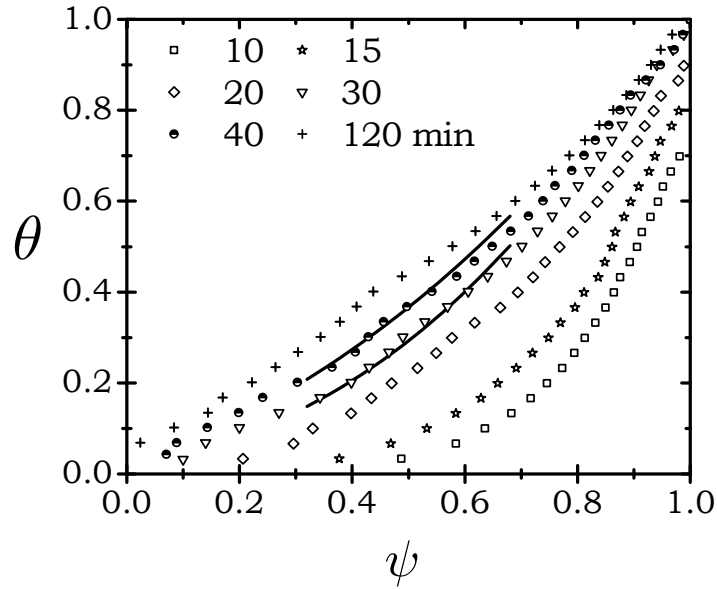

**Figure 4.** Evolution of normalized temperature ($\theta$) is plotted as a function of normalized distance from the bottom plate ($\psi$) for the interference patterns shown in Figure 3 (2.4 volume %). Symbols represent the experimental data obtained at different times (in minutes) while the lines are fits of the analytical solution (Equation (12)) at different times to the experimental data. In 90% sensitivity window in time and spatial domain the coefficient of determination ($R^2$) for the fits for 30 and 40 min data are 0.9826 and 0.9645 respectively.

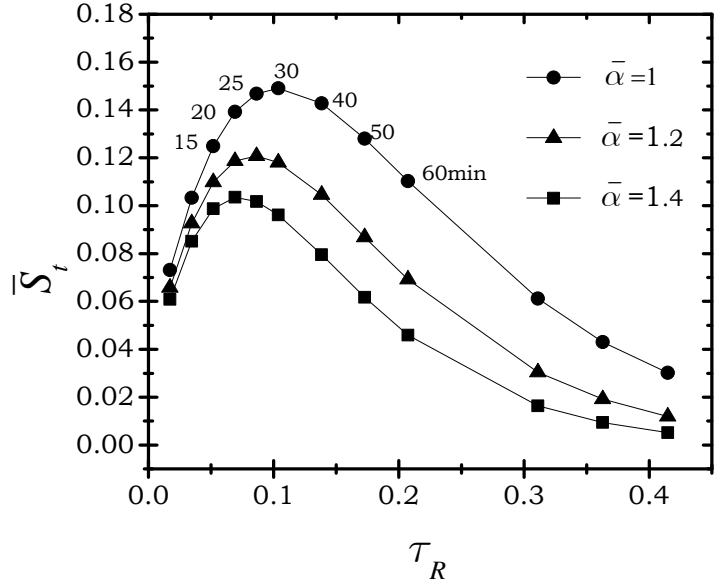

(a)

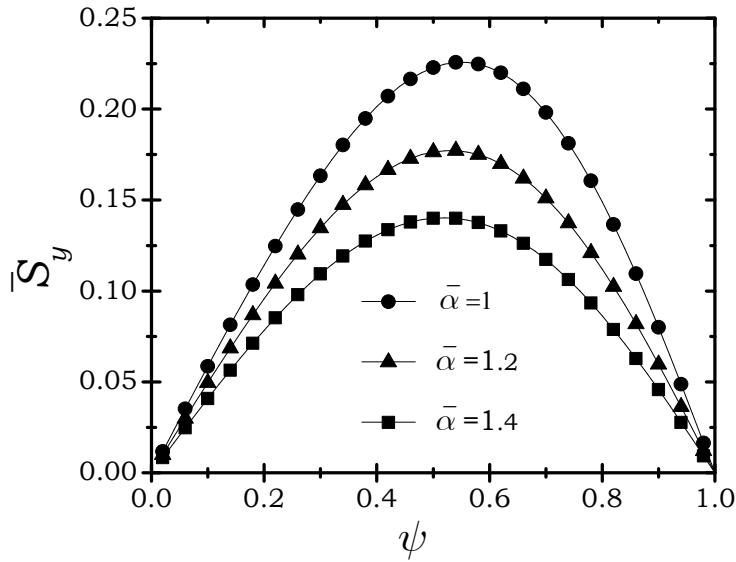

(b)

**Figure 5.** Normalized sensitivities $\bar{S}_t$ and $\bar{S}_y$ are plotted as a function of dimensionless time (a), and dimensionless distance from the bottom plate (b) respectively. In figure 4 we fit only that part of the data which lies within the 90 % of the window of the maximum value.

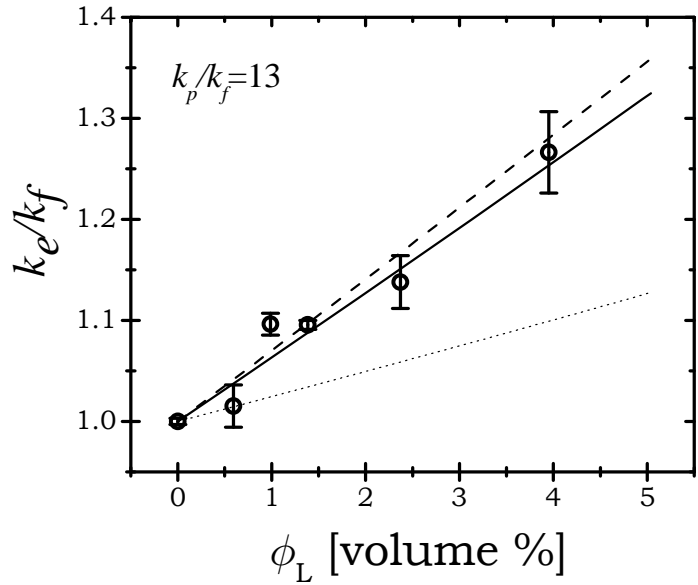

**Figure 6.** Thermal conductivity as a function of concentration of Laponite JS. Symbols represent the experimental data, solid line represents analytical solution due to Fricke (Equations 4 to 7), while dashed line represents analytical expression due to Hamilton and Crosser (Equations 4, 8 and 9) for $r_p=1/25$. Dotted line is a solution of Maxwell equation (Equation 4) for suspension of spherical particles ($n=3$). For all the cases conductivity ratio is: $k_p/k_f=13$.